# Quantum Objects

**Alireza Mansouri[*], Mehdi Golshani[†], Amir Ehsan Karbasizadeh[‡]**

**Abstract:**

In this paper, we propose an alternative interpretation of the quantum state vector, which, by considering *temporal parts* for physical objects, aims to provide a more comprehensible account of the measurement problem in quantum mechanics. We examine the capacity of this interpretation to explain three measurement problems: the problem of outcome, the problem of statistics and the problem of effect. We argue that this interpretation of the state vector while providing a satisfactory account and being as rationally plausible as its rivals for the measurement problem, reveals yet another limitation of our perceptual experience: our inability to perceive *unsharp reality*.

## 1. Introduction

Metaphysical doctrines, though, are not usually as criticisable as scientific theories, but sometimes two different metaphysical views offer two interpretations of a body of known facts, one of which is compatible with our temporarily best scientific theory. So, the metaphysics behind the incompatible one loses its interpretive power and is abandoned. This is how some scientific problems are relevant to metaphysics.[1]

The so-called measurement problem, from the inception of Quantum Mechanics (QM), has been one of the most controversial issues, and there is no general agreement among both physicists and philosophers of physics, even on the problem *itself*. Despite the fact that some authors, like Bub (1974:140), have taken the measurement problem as a "psudo-problem" arising from "a misunderstanding of von Neumann's problem", others have suggested various models and interpretations, each of which is based on a specific construal of the problem with a different focus, to solve it. According to Albert's formulation of the problem, for instance, "The dynamics and the postulate of collapse are flatly in contradiction with one another (just as we had feared they might be); and the postulate of collapse seems to be right about what happens when we make measurements, and the dynamics seems to be bizarrely wrong about what happens when we make measurements; and yet the dynamics seems to be right about what happens whenever we aren't making measurements." (1992:79). To be more precise, as

[*] Department of philosophy of science, Institute for Humanities and Cultural Studies, Iran; email: a_mansourius@yahoo.com

[†] Department of physics, Sharif University of Technology, Iran; email: golshani@sharif.edu

[‡] Department of philosophy of science, Iranian Institute of Philosophy, Iran; email: karbasizadeh@irip.ir

[1] See Agassi (1964)

Maudlin (1995) explained in his paper, entitled "Three Measurement Problems", the following three claims about QM are mutually inconsistent:

A) The wave-function of a system is *complete*.

B) The wave-function always evolves in accord with a linear dynamical equation.

C) Measurements of, e.g., the spin of an electron always (or at least usually) have determinate outcomes, i.e., at the end of the measurement the measuring device is either in a state which indicates spin up (and not down) or spin down (and not up).

To see the inconsistency of these three claims, consider, For example, a good z-spin measuring device, which has a ready state ($|\text{ready}\rangle$) and two indicator states (call them "UP" and "DOWN. The device is so constructed that if it is in its *ready state* and a z-spin *up* electron $|+z\rangle$ is fed in, it will evolve, with certainty, into the "UP" state $|"UP"\rangle$, and if a z-spin down electron $|-z\rangle$ is fed in, it will evolve, with certainty, into the "DOWN" state $|"DOWN"\rangle$. Using obvious notation:

$$|+z\rangle|ready\rangle \rightarrow |+z\rangle|"UP"\rangle, \text{ and}$$

$$|-z\rangle|ready\rangle \rightarrow |-z\rangle|"DOWN"\rangle$$

If we feed in this device an electron in an eigenstate of x-spin rather than z-spin, since

$$|+x\rangle \rightarrow 1/\sqrt{2}|+z\rangle + 1/\sqrt{2}|-z\rangle$$

then according to the claim B and linearity of the evolution, the initial state must evolve into

$$(1/\sqrt{2}|+z\rangle + 1/\sqrt{2}|-z\rangle)|ready\rangle.$$

But the question is what kind of state of the measuring device this represents:

$$|S\rangle = 1/\sqrt{2}(|+z\rangle|"UP"\rangle + |-z\rangle|"DOWN"\rangle)$$

If A is correct, and the wave-function is *complete,* then this wave-function must specify, directly or indirectly, every physical fact about the measuring device. But, simply by symmetry, it seems that this wavefunction cannot possibly describe a measuring device in the "UP" but not "DOWN" state or in the "DOWN" but not "UP" state. Since "UP" and "'DOWN" enter symmetrically into the final state, by what argument could one attempt to show that this device is, in fact, in exactly one of the two indicator states?

So if A and B are correct, C must be wrong. If A and B are correct, z-spin measurements carried out on electrons in x-spin eigenstates will simply fail to have *determinate* outcomes. This seems to fly in the face of Born's rule, which says that such measurements should have a 50% chance of coming out "UP" and a 50% chance of coming out "DOWN, simply because occurrence of outcomes is taken for granted within such a statistical framework.

To resolve the contradiction, people have pursued different strategies. In some theories, such as GRW, dynamics have been modified to a nonlinear one, while in others, like Bohm's model, it has been postulated to extend beyond what is represented in the wavefunction - namely, *hidden variables*. In yet another strategy, Everett-like theories, which abandon the very collapse postulate, have extended their ontology to include many worlds, many minds, and so on. Every proposal for tackling the measurement problem should have enough resources to give a satisfactory account of these three problems: (Maudlin 1995):

**The problem of outcomes.** In QM, we encounter some weird superposition states, which are linear combinations of pure states. The problem associated with these superposition states, which are themselves pure and independent, is that after the measurement process, the system randomly chooses one of the states in the superposition and eventually ends up with a defini*te* possible value for the associated observable quantity. In fact, "It is just the usual understanding of the *Schrödinger cat problem* laid out in a formally exact way. Oddly enough, when so laid out, it becomes immediately evident that a fair amount of the work in the foundations of quantum theory misses the mark. The most widespread misunderstanding arises from the claim that the measurement problem has to do with *superpositions* versus *mixed states*." (Maudlin 1995:9)

**The problem of statistics.** QM suggests probabilistic results satisfying Born's rule. Measurement situations, starting with the same initial wave functions, could lead to different outcomes. The probability of obtaining each outcome is in accord with Born's rule. The problem arises when wave functions evolve according to linear dynamics. Why do similar wave functions end up with different outcomes, in accordance with Born's rule?

**The problem of effects.** This problem arises from the fact that, according to the standard interpretation, measurement processes involve collapse, which changes the state of the system and thus influences its future development. The state of the quantum mechanical system after measurement should be an eigenstate corresponding to the measurement outcome. The primary motivation for considering such a condition is that the measurement is *repeatable*. To attain repeatability, the system after measurement should be in an eigenstate corresponding to the relevant observable, so that the second measurement, which is done *immediately*[1] after the first, yields with probability 1 the previous result.

Facing these difficulties, many physicists and philosophers have called for some new satisfactory conceptual schemes and categories. Moving in this direction, we propose a new interpretation of the quantum vector state, offering a satisfactory account of the measurement problem that we believe is as *rationally plausible as* its rivals.

The paper is structured as follows: Section 2 presents a new perspective on observation and measurement, which the SI is based on. Section 3 provides the Sigma interpretation of the quantum vector state. We address the three measurement problems (outcome, statistics, and effect), showing how SI resolves each in Sections 4 to 6. Section seven is the conclusion.

[1] As for 'immediately' condition, it is only important that, at the time of the second measurement, Schrödinger evolution does not change state of the system.

## 2. Observation and Measurement in the New Picture

According to three-dimensionalism (3D), objects persist through time by "enduring," that is, by being wholly present at all times at which they exist. In this account, the system cannot be in both states A and B simultaneously; observing it always yields a determinate value for the relevant quantity. Based on such a picture and presupposition, if there is a theory for describing a physical reality that attributes the superposition of states A and B to a system, then we would be in trouble interpreting how the system could be in such a weird state. Therefore, we might consider this description of physical reality incomplete (i.e., there should be additional parameters and variables to specify whether the system is in state A or B, or the theory is simply incorrect).

There is, however, yet another way out of this dilemma by suggesting an alternative picture for the evolution and observation of the system, that is claimed to be consistent with superposition states. To address this, we emphasise that QM needs to introduce a new conceptual framework to overcome its challenges. By appealing to some advantages of the new interpretation of the vector state, we will show that solving the measurement problem in QM requires a four-dimensional ontology of objects, according to which objects persist four-dimensionally through time with temporal parts, that is, by being temporally, as well as spatially, extended.[1]

Based on this interpretation, the physical object coincides with time and, like time, it is a *continuum*[2], which has *flowing*[3] *temporal parts*: the outcome of measuring an observable of the system, *at this moment*, is always something *other than* the outcome we would obtain in another measurement event at some other time, past or future. At each time slice, what is observed is a *new stage* (or a temporal part) of a flowing, unsharp observable (or a four-dimensional object) at every moment. In other words, as time passes, the old stage of the preceding observable is annihilated and the new stage of it comes to exist in its place, and this renewal happens *continuously* at every moment. 'Continuous', here, implies that the mixture is an *improper*, as opposed to be a *proper* mixture, of instant sharp observables in an interval of time. The *unsharp observable* is continuous and unified, much like a continuous spectrum; therefore, it is actually a *unified continuous reality*. So, objects are thought to have instantaneous temporal parts ('time-slices'), which do not themselves persist through time. This mode of reality, in its totality, cannot *wholly* exist simultaneously and cannot be *wholly* found at every moment, but its *totality* exists in its whole lifetime. For an intuitive picture, consider the following scenario.

---

[1] The original idea of this paper was under the influence of the ontology of Sadra's philosophy (1571–1641), especially, Obudyyat's (2006/1385). This idea was completed by recent attempts in four-dimensionalism ontology and its applications in physics. In this regard, especially, Balashov (1999) was provocative.

[2] The main feature of a *continuum* is that it is hypothetically divisible. That is, for every *moment* of time, there is always an instant*, after* that *moment,* which can be, in relation to that moment, considered as *future,* and an instant, *before* it, which can be, in relation to that moment, considered as *past*.

[3] A continuum could be *persistent* or *flowing*. *Persistent* continuum, like line, area and volume, is a continuum whose parts exist altogether. However, none of the two hypothetical parts of a *flowing* continuum, like time, exist together.

Imagine an observer who is an inhabitant of a flatland, a two-dimensional world populated by beings having only two spatial dimensions and wholly confined to a surface. For simplicity, let this surface be a plane. Now, suppose they are presented with an object, such as a pyramid. As shown in the following picture, the flatlanders perceive reality only in a two-dimensional plane and have no perception of the third dimension; in other words, their situation renders them incapable of perceiving the third dimension. They are ignorant of what there *is,* in the other dimension external to his plane.

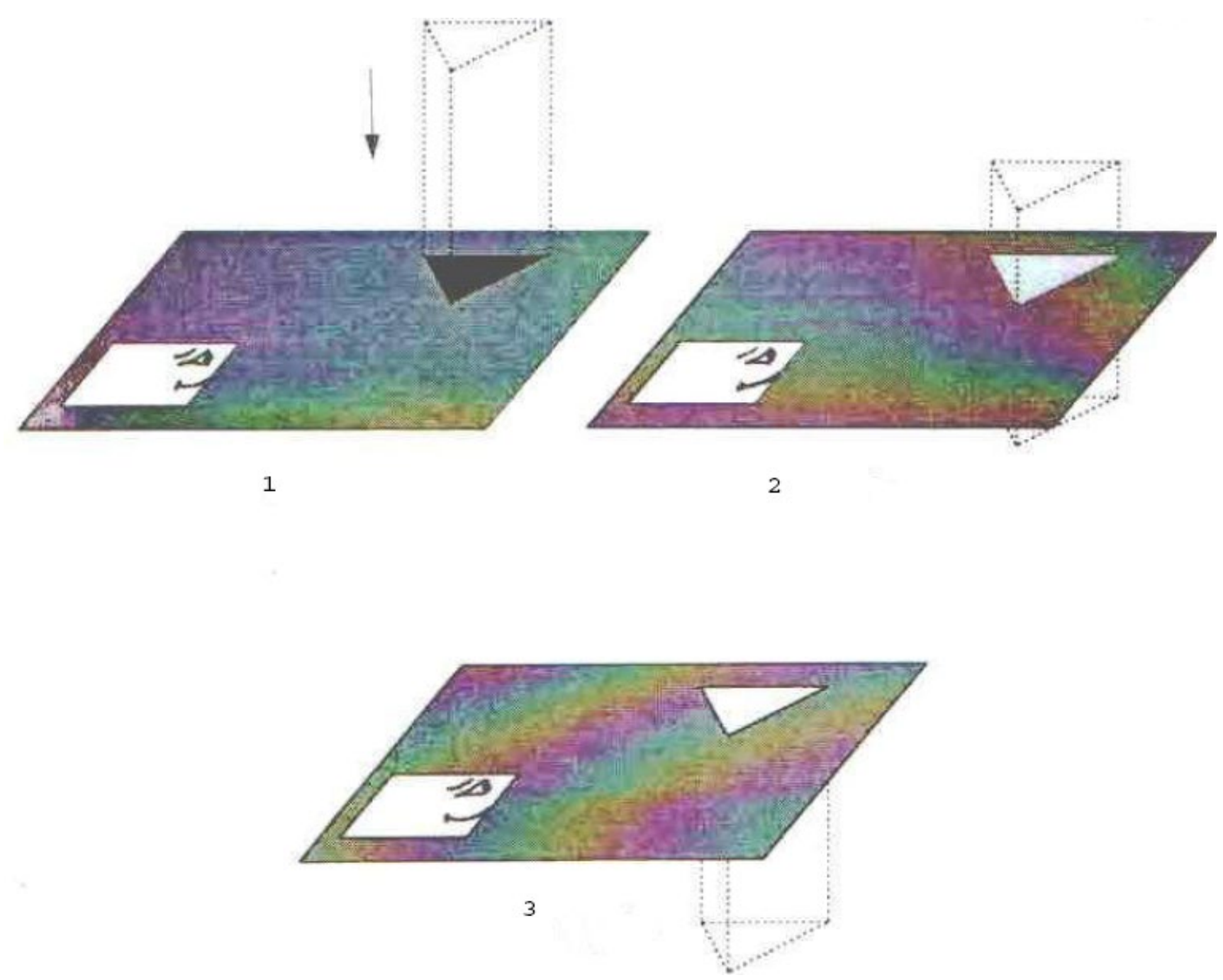

Now imagine a *purely* and uniformly coloured black pyramid with a triangular base; this pyramid is in a *pure state*. Before the pyramid enters the plane, there is nothing for the flatlander to see. However, when it enters the observer's world and the plane intersects with the pyramid, the observer sees only a cross-section of the pyramid, i.e., a black triangle. This triangle persists in existence for the flatlander by enduring through time intervals in which the pyramid passes the plane. As a matter of fact, the flatlander perceive two-dimensional slices.

Now imagine another observer who perceives the third dimension. They would realise how the observer in the plane could see the pyramid: the flatlander *falsely* thinks and report that a triangle comes to exist at a moment in time, and after passing some time, it is annihilated. However, the situation for the second observer is radically different, as no triangle is being observed. Therefore, from the second observer's point of view, what the flatlander perceives is a ‘hypothetical’ cross-section of the pyramid with the plane. Moreover, the flatlander, due to the pure black colour of the pyramid, has the illusion that there is only one and the same triangle, which persists over time and has never undergone change or evolution. This is an *illusion*, however, from the second observer's point of view, which is outside the plane. They know that the flatlander has never seen the same triangle in any two moments. Still, because these triangles are in complete similarity with each other, it *appears* to them as only *one* persistent triangle. They have observed different triangles, which are temporal parts of a 3D object, i.e. the pyramid, at different instants of time.

The two observers both provide a true description of their observations, but by considering the additional dimension, the description of the second observer is more comprehensive. One might ask whether the flatlander could reach to the second observer's picture at all. Certainly, he could, but not by direct observation. He could discover that picture through imagination, theorising, or mathematical abstraction.

Therefore, according to our account of objects, if we accept that objects and physical systems have temporal parts, it follows that the sentient observer always perceives only a part or a stage of the unsharp observable of a flowing system since mental experience occurs momentarily. As it was said before, the flatlander never sees the three-dimensional pyramid and only sees a cross-section of it. In the same vein, observers in a measurement process never see the entirety of the flowing system, which coincides with time. Instead, in every measurement and observation, the observer sees only the hypothetical cross-section or a stage of it, which is in its plane. In our example, although it appears to the observer that he sees the same triangle, in fact, they see a new triangle at every moment of time. Similarly, our observer in the measurement process has the same feelings regarding the observing system. That is, it *appears* to them that they always see *the same* system, while the system is always becoming new. He is unable to feel the evolution of the new system for the same reason that applies to the flatlander. Those triangles that the flatlander sees are not discrete from each other, rather, they are all 'hypothetical' cross-sections of a continuous persistent system. The determinate values observed by the observer are *not* 'discrete' and 'isolated' from each other. It is also not the case that the observer measures the observable independently of what they have observed or what they will observe. All of them are hypothetical continuous cross-sections (or temporal parts or temporal stages) of what is called a *flowing* system.

In the flowing picture, therefore, the system, with all its unsharp properties, evolves altogether; however, the observer thinks that the system (or other properties of the system) is at rest and does not evolve. It appears to him that, in every moment, the system exists with all of its sharp properties and sharp observables. After that moment they are all annihilated and the system with all of its observables are simultaneously renewed again, and so on. This happens, however, in a way that these *becoming* sharp realities, in sum, create a *flowing* object, without any *discontinuity* between *them*. Based on this picture of reality, we give an interpretation of the quantum vector state.

## 3. Interpretation of the Quantum Vector State

Unless we change our interpretation, it would be meaningless to say that an electron, which is in a superposition state, is in several positions simultaneously. Many interpretations have been suggested for such weird states and to explain how they could be reconciled with our definite experiences. In the many-worlds interpretation, for example, there are several copies of electrons in different positions in different worlds. This, however, is not the only way out. The *Sigma interpretation*, based on the picture of observation and evolution presented in the previous section, is another way of attempting to do so, similar to the many-worlds interpretation, without the collapse.

SI makes two physical claims: I) The state vector of a system completely describes the physical reality of that system. II) The state vector evolves in accordance with Schrödinger's equation. The problem, as we have seen, is how to reconcile II with the appearances, especially with the definiteness of our experience. According to this interpretation, if we understand the state of the system in a certain way, we will see that the collapse and the probability interpretation emerge as features of the theory. The state of the system $|S\rangle$ can be represented as a superposition thus:

$$|S\rangle = c_1|a_1\rangle + c_2|a_2\rangle \dots$$

where the $a_i$ constitute a complete set of basis vectors that span the Hilbert space for the whole system. Our suggestion is that each of the components of the superposition be thought of as representing the density of the local state, in which the temporal parts of the flowing system are created or renewed.

How can the flowing object save appearances? Suppose that an observer M measures the x-spin of an electron e in an eigenstate of z spin. Let +z be the state z-spin "up", +x (-x) be the state x-spin "up" ("down"), $B_o$ be the state of the observer before making the measurement, and +B (-B) be the state of M observing that x spin is up (down). Before the measurement, the state of the system is $|\mathrm{Bo}\rangle|+z\rangle$. After the measurement, the state of the system (observer+electron) is

$$1/\sqrt{2}(|+B\rangle|+x\rangle + \ |-B\rangle|-x\rangle)$$

According to the SI, the flowing electron has been renewing and has been creating, *at every moment*, in “spin-up” OR “Spin-down”. So, when M measures the spin of the electron, M observes definitely “spin-up” OR “spin-down”. Each time, she has a perfectly definite experience corresponding to the perfectly definite stage (temporal part) of the electron.

So, we take the wave function as a description of a flowing system. In the flowing picture, *at every moment*, only *a cross-section* or a stage of the physical system, not *the whole* physical system, could be found. Moreover, the observer finds every cross-section or stage only at *a* moment, and an individual stage can't be found at every moment[1].

According to this interpretation, the state vector can be considered as a description of a single, isolated system. Whereas the wave function describes the system as a *flowing* object, probability takes on a new meaning; obtaining a determinate outcome in a measurement process is related to the statistical distribution of potential states and stages of the flowing system, which are reflected in the superposition state. The system, therefore, is always in motion and evolution during its entire lifetime, and its becomings occur with a new value for the observable throughout its entire lifetime. The probability attributed to the renewals, however, is governed by a specific distribution, i.e. according to Born's rule[1], and, of course, at every moment of time, the distribution of temporal parts of the "quantum object", described by the quantum state, evolves according to the time-dependent Schrödinger equation.

[1] As in Everett's (1973:61) interpretation, this result is independent of the *size* of the object, and remains true for objects of quite macroscopic dimensions. That is, they are also *flowing* reality, which are followed by the fluidity of their parts. So, Sigma interpretation, as Everett's (1973:87-8), gives a unified description of micro- and marco-objects.

von Neumann's interpretation of the state is *not* realistic and takes it merely as a *theoretical construct*. On the contrary, in the Sigma interpretation (SI), as in Everett's, state vector refers to a real state of affairs. From a realist stance regarding quantum states follows the reality of superposition states and their implicit indeterminacy. Indeed, according to the SI, the system is *principally* in a superposition state; however, when, according to the standard view, the system is in a pure eigenstate, the terms in the superposition are the same, even when the system is in a *pure state*, it should be interpreted that the system is in continuous renewals in 'states', which happen to be, in complete similarity. So if the state of the system is in the superposition of states corresponding to the black and white colour we should not interpret it as "being black and white" probability; rather, we should talk about the probability of "becoming" black or white. That is, the system is *becoming* black and white at every moment. Of course, the system, in its renewal and becoming, has sharp observable quantities, but these quantities do not have *distinguishable* values. All systems and states, therefore, even when they *appear* to be at rest, are in continuous renewal and becoming.[2] One may object that there is nothing to identify these renewal states as constituting the same individual's experience. In replying to this objection, it should be noted here, in passing, that since, in the SI, the reality of superpositions preserves the trans-temporal identity of systems, this perpetual renewal does not violate the *identity* of the physical system.

## 4. Problem of Outcome

If wave function is considered *complete*, then it must specify every physical fact about the composite system and the measuring device. But by what argument could one attempt to show that the composite system, in a superposition state, actually gives in exactly one determinate outcome? That is the *problem of outcome* in the measurement problem.

According to the mentioned account, and using Balashov's (1999) terminology, the relationship between the spatial properties of the pyramid and those perceived by flatlanders is a rather intimate, ancestral one. The 2D shapes are directly "inherited" from the 3D shapes, as they are constituted by one-dimensional sides that entirely belong to the two-dimensional edges of the original three-dimensional object. It is an *objective* feature of how Flatland as a whole is situated in the wider three-dimensional "superspace." The main discovery of the Flatland physicists is the discovery of this objective, ontological feature to which they have been led to by an argument to the best explanation. A physicist from Flatland could hypothesise that the triangle is a cross-section of a three-dimensional being and should be a pyramid situated in Flatland, as shown in the above Figure.[1]

[1] So, based on this interpretation, it is *not* compelling to take probability or chance as an *inherent* feature of the world. As in Ballentine's statistical approach, perusing hidden variables is legitimate. Above all, it is still necessary to provide an explanation for the specific distribution governs renewals.

[2] This is very similar to Bohm's *'holomovement'* concept. The idea that everything is in a state of flow or becoming (or what he calls the "universal flux"). For Bohm, wholeness is not a static oneness, but a dynamic wholeness-in-motion, in which everything moves together in an interconnected process. See (Bohm & Hiley 1993:355), (Pylkkanen 2007:25), and Bohm (1980:11)

In the relative-state interpretation, all possible measurement outcomes are considered equally real for a fictitious, non-existent (or global) "observer" outside the Universe who views the universal state vector, whereas for each observer inside the Universe, it is always one particular value of the measured observable that is real. In the same vein, within the Sigma Interpretation of the measurement process, the relationship between unsharp observables and determinate outcomes, as perceived by the observer, is a rather intimate and ancestral one. The determinate outcomes, which the observer perceives, are slices of a flowing, unsharp reality, which, from an *atemporal* observer's standpoint, *indistinguishably* sit beside each other. Because of the limited perceptual ability of the usual observer (that is because he cannot have a perceptual experience of the whole system at a moment), they can only see a determinate outcome. In other words, as any observer's experience happens at a moment of the continuum of time, and as the observer observes the system from a lower dimension and intersects with the flowing reality, only at a moment or at a point of flowing continuum, he inevitably sees a determinate outcome. After all, we cannot observe the total lifetime of an entity or even a part of it at a moment.[2] Accordingly, as in many-worlds interpretation, in which determinate outcomes in every world explain the *empirical content* of the theory, in the same manner, in the SI, temporal parts of the fluid object, which are *indistinguishably* determined at every moment, account for the *empirical content*.

In hidden variable theories, it is taken for granted that experience provides a 'complete' description of reality; since we experience *determinate* outcomes, we conclude that reality is completely *determined*. On the other hand, since, in quantum mechanics, we encounter superposition states, which imply indeterminism, we must eventually consider quantum mechanics an incomplete theory. After all, it could not have yielded a representation of determined reality (the reality we reach based on our sensory experience). Accordingly, there needs to be additional parameters to make it complete. According to the SI, however, contrary to what a hidden variable theorist believes, our experience does not provide a 'complete' description of reality, and the determined experience arises from our inability to perceive indeterminate superposition or *unsharp reality*. In fact, we encounter a situation which Duhem-Quine's thesis can describe. The SI *prefers* to take the *appearance* of 'determinate outcomes' as a *product of our mind*.[3] Although we are unable to have a perceptual experience of a

[1] Here, we have almost used Balashov's exposition of the scenario (1999).

[2] Contra Everett's theory in which all elements of superposition states represent *actual* realities which simultaneously exist, in Sigma interpretation, elements of superposition state represent *potential* realities which *briefly and indistinguishably* exist in a *flowing continuum* represented by the superposition state in its totality. This is followed by the fact that in Everett's theory, superposition state represents a persistent system, which exists through time by enduring, while Sigma theory interprets superposition state as a *flowing continuum*.

[3] According to von Neumann, as far as formalism of QM concerns, there is no *physical* collapse, and we encounter with superposition states. To obtain determinate outcomes, he believes in a *mental* collapse. On the other hand, for any determination in mental level, it seems that there should be a correspondent determination in *physical* level. Handling this problem, von Neumann suggests an *arbitrary* distinction between observed system and observer (von Neumann 1955:421), Becker (2004). Although this "arbitrariness" does not explain *why* and *how* the collapse occurs, the SI claims to explain it by suggesting a new *ontology and conceptual scheme*. It

superposition state, we can reach those indeterminate superpositions by theorising or mathematical abstraction, as we reach the heliocentric picture of the solar system through theorising rather than *direct* observation or sense experience.

By these considerations, we adopt a realist position regarding superposition states, viewing them as describing a mode of existence of reality and its temporal parts. These temporal parts have a *brief* and *potential* reality in superpositions, *not* in the sense that those states *do not* exist outside of the superposition, nor are they unreal.[1] They are *real* but not in an *independent* and *distinguishable* way. In the SI, superposition states of a dead and alive cat represent the unsharp reality of both the dead cat state and the alive cat state. The superposition state of a cat describes Schrödinger's cat from an atemporal perspective as a flowing system with an *unsharp reality*. Therefore, the state of the cat is potentially both alive and dead. Not in the sense that the cat is *not* alive (or dead) at this moment, but can be alive (or dead) at another moment. However, in the sense that the cat's flowing reality can be found alive (or dead) by cutting the flowing reality of the cat, i.e., by measuring it and by the *participation of mind and consciousness* in the process.[2,3] It should be noted that, in the Sigma model, obtaining a determinate outcome is the result of the *special metaphysical relation* of the observer as regards to the system, *nay* the special role of the measurement. That makes the state of the observer depend upon the system. It depends on which plane it cuts the flowing system. This *relativity* is not also a newborn feature. Copenhagen and Everett's interpretations also share this feature.[4] In the Sigma model, however, there is an explanation for it. The dependency of

---

should be noted that the '*product of our mind'* does *not* mean that the determinate outcomes are *not real*. They are *briefly and indistinguishably* exist in a *flowing continuum* (see footnote 2, p. 9).

[1] Heisenberg (1958:53), used the idea of Aristotelian 'potentia' in his later work on the foundations of the quantum theory. He goes beyond the usual interpretation of the pure quantum state as the catalogue of all actual properties—those with probability equal to one—of an individual system in that he considers $\psi$ as the catalogue of the potentialities of all possible (sharp) properties *Q* of the system, quantified by the probabilities $p_{\psi}(Q) = <|\psi|Q\psi>$. The notion of *actualization of potentialities*, then, makes it possible for one to "say that the transition from the 'possible' to the 'actual' takes place as soon as the interaction between the object and the measuring device, and thereby with the rest of the world, has come into play; it is not connected with the act of registration of the result in the mind of the observer." (1958: 54-55). Potentiality has been also articulated by Shimony (1993:179) as a *modality of existence* of physical systems. But Heisenberg and Shimony's conception of "potentiality" is not in *temporal parts* context.

[2] So, Copenhagen interpretation is right, *in a sense,* when it says that we have a *determinate* outcome *just after measurement*, but the importance lies in *explaining* it. the meaning of this claim is *criticizable* and rational only in a metaphysical framework, not by denying metaphysics and ontological theorizing.

[3] In another project the concept of *unsharp quantum reality*, though *ontologically* different from ours, has been also suggested, and a particular way of formalizing it was explored by Busch *et. al*, (1996) to explain quantum indeterminacy and measurement problem. However since, in this approach, there is no room for the role of *mind*, it requires a stochastic modification of the unitary time evolution postulate for quantum dynamics. See Busch & Jaeger (2011).

[4] Bohr criticized EPR's 'physical reality criterion'. According to Bohr, this criterion is essentially ambiguous because it is proposed without considering experimental *arrangement* (Bohr 1935:697). According to the *fundamental relativity of quantum mechanical states,* in Everett's *relative* state interpretation, likewise, we speak of observer's state *relative* to the system's state.

subsystems on other parts of the composite system, moreover, explains both indeterminism and continuity, and it means that the whole has an ontological priority over its parts.[1]

### 4.1. Sharp Reality and Unsharp Reality

As mentioned above, in the SI, a physical system has two modes of reality: one mode is unsharp, and the other is determinate or *sharp*. What we measure as a sentient observer is its determinate feature, which *appears* to the mind as a sharp reality, and what acts in nature without the contribution of the mind is the *unsharp* reality. For example, +1/2 and -1/2 in the z direction at every moment are a sharp reality of the spin of an electron. In this sense, spin exists in a sharp or determinate mode of reality. On the contrary, different states in the superposition correspond to different values but in a *potential* and *indistinguishable* way. So, spin has another kind of reality, which is *unsharp*. It has both +1/2 and -1/2 values, but in a *brief* and indistinguishable way, of course, flowing.

According to the SI, the superposition state of a flowing object refers to a genuine, simple, and *unified* reality. We should *not* think of it as representing the *sum of*, or a real mixture of, multitudes of individual realities attached together. This explains why, in QM, we ought to differentiate between *superposition states* and *mixed states* to distinguish between proper mixtures and improper mixtures. Although the superposition is a reality, which is genuine and simple, it does *potentially, indistinguishably, and briefly* include parts other than what has been determinately observed by the observer. It follows that the reality of the superposition does not imply a mere *multiplicity* and *plurality*, which ultimately leads to the difficulty of how an object can be *simultaneously* in multiple exclusive states.

Moreover, the superposition state provides a more complete description of reality than what is given by our observations and measurements as a determinate outcome and as what appears to us as mental reality. The superposition state's description of physical reality is a description of a higher order or one that considers a higher dimension regarding the determinate observed outcomes. This matter, accordingly, leads us to the conclusion that the reality of superposition states and indistinguishable realities should be considered as the origin of the determination and definiteness of outcomes of measured quantities, not as a denial of it. If we were to consider the unsharp reality from the lower dimension, it would appear to us as a determinate outcome, *specific*, *definite*, or *distinguishable*.[2] However, if we were to consider it from a higher dimension, the 'superposition' would still be superposition.

[1] This *wholeness* in Bohm's model, that is the fact that " … not only the inter-relationships of the parts, but also their very existence is seen to flow out of the law of the whole." (1971:viii), is explained by dependency of quantum potential on the state of the whole system.

[2] Attempts to suggest new interpretations show that we have to *extend* our ontology and provide new *conceptual schemes* to understand the measurement process in QM. In Everett-like interpretations, this extension appears as additional worlds or minds, and in Bohm's model it appears as additional variables and a new potential with some weird properties. In Bohm's model, which denies superposition states, *quantum potential* is responsible for all of those strange behaviors which result from superposition states (Riggs 2009:127).

Since superposition states are considered for *certain* observable and quantity, they are written in *certain* basis of space[1]. These are superpositions of pure states, each corresponding to a specific value of the relevant observable. As for the pyramid and triangle picture, there is an ancestral relation between pure states and the superposition that they create. This relation, since it has an ontological meaning in SI, explains how we encounter the *two modes of realisation* of *the same* physical quantity, which are represented in the *same* space.

Denying the reality of the superposition states originates from the presupposition that systems could only be *determinately* and *definitely* realised. But there is no compelling reason for this presupposition. We could consider systems as *four-dimensional* objects, though indeterminate from a four-dimensional point of view, which appears determinate from a three-dimensional standpoint. This does not mean that unsharp reality, in its totality, exists at every moment of its lifespan. Since the total continuum of the unsharp observable, corresponding to the superposition state, is not realised at every moment (it is a flowing reality), *it is not possible* at all for the observer to see the whole continuum of *unsharp observable* at *a* "moment", but he sees its total continuum in its total lifetime.[2] The determinate value we see for an observable quantity at every moment is only a hypothetical cross-section of the flowing continuum of the unsharp observable corresponding to the superposition state. In other words, the observer is right in describing and reporting his experience, but he is wrong in *interpreting* his determinate experience as the *ultimate* and *complete* reality.[3] From the atemporal observer's point of view, or based on the quantum description of reality, the system is not in a determinate state but rather in a superposition of states.

### 4.2. A Critique Against the Existence of Superposition

One might object that since the pointer of a measuring device is a pretty classical object, and the different values obtained in measurement are macroscopically *distinguishable*, it makes no sense to say that it has no definite position at any instant of time.[1] This type of objection considers the relationship between *distinguishable* values of a quantity and its *reality*. In the SI, however, it should be emphasised again there is no distinguishable position for the system before the measurement, and one cannot deduce *nonexistence* from *indistinguishability*.

[1] Furthermore, while Dirac developed the general principles of quantum theory, he gave no *preferred* role to picturing processes explicitly as occurring in *space*, the democratic equality between different points of view was maintained in the new dynamics that resulted. All observables, and their corresponding eigenstates, had equal status as far as fundamental theory was concerned. The physicists express this conviction by saying that there is no 'preferred basis' (a special set of states, corresponding to a special set of observables, that are of unique significance). Wrestling with the measurement problem raised in the minds of some *the problem of preferred basis*. But, in the SI, this is by no means a fundamental necessity. Because of the continuous *total* renewal of the object, at every moment of time, and since our object, renewed with *all of its properties*, consistent with Dirac's formulation, there is no need to an *objective* preferred basis. The *so-called* preferred basis is as a result of what we (as observer) *prefer* to choose for measurement.

[2] While there is no plausible account for *unobservability* of Everett's branches (or other worlds in many-worlds interpretation) (Barrett 1999:158-9), there is an explanation for *unobservability* of all states or outcomes at *a moment*: all physical systems and entities are in correspondent with *continuum of time*, and because of the flowing nature of them, as in the stage theory, they persist through time by *perduring*, *not enduring.* I used the term 'perdure' to hold the basic idea that persistence is much like spatial extension. These systems, therefore, are four-dimensional objects with temporal parts, but we experience only a brief temporal parts or 'stages' of those four-dimensional objects.

[3] Comparing with the bare theory makes this more clear. According to the bare theory, when a system is in superposition state, the observer would *falsely* report that he got some determinate result, because, as a matter of fact, there is no determinate state (Barrett 1999:96). But according to the Sigma interpretation, the observer has really a determinate experience, he *mistakes* when he thinks what he observes at every moment is *the whole* of the system.

To be more intuitive, consider the real numbers axis, which consists of numbers *innumerably* and *indistinguishably* ordered along the axis. This does not mean they do not exist. They exist, but *not* in a distinguishable way. Here, we emphasise again that the meaning of the potentiality of states in a superposition state is not that the system, at present, has no definite state and could have it rather, it means the system *indistinguishably* takes the values corresponding to those states, just as real numbers axis, which is real and genuine, *innumerably* and *indistinguishably* includes numbers. When it is said that numbers *indistinguishably* and *innumerably* exist, it does not mean that the axis is devoid of any numbers. It *innumerably* consists of all numbers. It is possible to distinguish and specify a definite number in the real numbers axis only by cutting the real numbers continuum. In the same vein, only by cutting the superposition state, i.e., by performing a measurement on the system that is in a superposition state, can we reach a determinate and distinguishable state. Consequently, a determinate outcome corresponding to the pure state for the relevant observable could be obtained.

On might object that individuality is based on *definiteness* and *distinguishability* from others. When we accept that a system is in a superposition state and is not distinguishably in any parts of the superposition, and when we assume that its definiteness and determination can only be reached through cutting the superposition in hypothetical moments or points of time, it follows that those parts of superposition are 'hypothetical' and 'unreal'. That is, the system lacks any values for the quantity. In other words, it seems, since before measurement, the system is in a superposition state, it lacks any definite value for that quantity, which, in turn, corroborates Heisenberg's assertion: there is *no* observable *before* a measurement[2]. This is the same conclusion that Einstein, rightly, but based on a certain metaphysical framework, refused to accept by asking the question: Do you mean that when you do not look at the moon, it does not exist?

A system in a superposition state, before a measurement, actually has values for the relevant observables. However, it is essential to note that these "actual" values or individuals are for the flowing and persisting observable quantity, which extends four-dimensionally through time. The above objection had some force if the assumption that there is no observable before the measurement was plausible. But in the light of our account and by taking quantities of the system as *flowing* properties, then the system undoubtedly consists of values for the observables.

---

[1] See Ballentine (1970:368-9) for this objection.

[2] Heisenberg's famous dictum in his (1927) *Uncertainty Principle Paper* was that "I believe that the existence of the classical "path" can be pregnantly formulated as follows: The "path" comes into existence only when we observe it."

## 5. Problems of Statistics

Quantum Mechanical predictions are probabilistic, based on the Born's rule. The usual application of Born's rule states that the outcome of an experiment is not predetermined, and there is, in fact, a certain chance of obtaining a possible outcome. One might ask how it is possible that the complete wave function always evolves in accordance with a deterministic dynamical equation (i.e., the Schrödinger equation), but in measurement situations described by *identical* initial wave functions, measurement sometimes yields *different* outcomes. The probability of each possible outcome is given (at least approximately) by Born's rule?

To account for the problem of statistics, consider the case of pyramid, plane and flatlander again. Suppose, further, that the pyramid has been coloured in black and white. These colours are *exclusively* and *randomly* distributed in the pyramids so that cutting every cross-section of it, gives us a black or white cross-section. Although randomly distributed, we can see, in sum, that half of the pyramid is white and the other half is black. Therefore, if we cut a cross-section of it, there is a 50% chance of getting white and a 50% chance of getting black.

What does the flatlander observe in the plane? Since he is unaware of the third dimension, he observes a triangle that randomly changes from black to white. But if this pyramid passes through the plane, times and again, he could eventually conclude that, in sum or in the long run, the triangle, half of its lifetime, becomes white, and half of its lifetime becomes black.

On the other hand, from the external observer's perspective, there is no *real* triangle. However, there are triangles that are cross-sections of a pyramid, coloured in black and white, with a specific distribution. It *appears* to the flatlander that there is a triangle becoming white and black. Moreover, for the external observer, the pyramid, in its entirety, is neither black nor white, and it is only after cutting the pyramid that we can say the cross-section is white or black. A flatlander's mind, randomly but based on a specific distribution, experiences a specific and determinate colour. The evolution of the flowing object wave function, or, in Everett's words, the evolution of the global wave function, is causal and deterministic. However, since the observer's mind experiences the randomly changing black and white stages of the flowing object, based on a specific distribution, the evolution of the flatlander's mental state is probabilistic and random. Accordingly, like Everett's model in the SI, although wave mechanics (Schrödinger equation) is deterministic, it gives rise to probabilistic appearances at the *mental level* in such a way that the relative frequency of determinate outcomes is square of coefficients of superposition elements (Everett 1973:78), (Barrett 1999:79-81). However, while Everett's theory does not explain why the observer has such mental experiences, the SI offers an intelligible account of it.

The SI accepts the common sense and intuitive idea that no individual system can be in different states at the same moment. However, it is possible to be in different states at *different* moments. Wave functions, in this interpretation, describe a flowing system that evolves according to the Schrödinger equation. The system has a mode of existence or is in a state of affairs, known as the *superposition*, which originates from the fact that the system does not persist by enduring through time; i.e., it is not wholly present at every moment of time at which it exists.

These renewals, as in the statistical interpretation, are governed by Born's rule. We use the *ensemble* concept to define probabilities because attributing a probability to every new state of happenings poses conceptual difficulties. The technical reason is the same as in statistical mechanics. As Albert (2000:63) put it:

> "it has to do with the fact that the sort of information we can actually *have* about physical systems—the sort that we can *get* (that is) by *measuring*—is invariably compatible with a *continuous infinity* of the system's *microconditions.* This follows from the fact that the totality of the possible microconditions of any Newtonian system invariably has the cardinality of the continuum and that the accuracies of the measurements that we are able to perform are invariably *infinite.* But the only way of assigning equal probability to all of those conditions at the time in question will be by assigning each and every one of them the probability *zero.* And *that* will, of course, tell us *nothing whatsoever* about how to make our *predictions*."[1]

The dispersion of the measuring device pointer in the SI, expresses possible measurement outcomes, and as far as the observation is concerned, there is no difference between the Sigma and the statistical interpretation in this regard. *Experimentally*, it makes no difference to say that we have a flowing system with continual renewals and a specific distribution or we have different systems that are similarly prepared. When we measure them, we would find a specific distribution for the positions. The difference lies at the *ontological level*.

The statistical interpretation of the state vector cannot be considered a description of an *individual* system. Ballentine rightly maintains that it makes no sense to say that the pointer of a macroscopic system has not a determinate position (Ballentine 1970:371). However, it makes no sense in a *specific metaphysical framework*, and it might be possible and meaningful in a different metaphysical framework.

Contrary to the ontology of statistical interpretation, as we consider quantum systems as flowing and four-dimensional objects, SI takes dispersion as a property of the individual systems. These objects do not persist by enduring through time. That is, the system could not *distinguish all the possibilities and could not take all the possible positions at every moment. In contrast, quantum systems indistinguishably and potentially take all those possibilities. Quantum objects persist four-dimensionally through time, and the observer's measurement mentally creates a distinguishable, determinate position for the*m.

Everett also makes a difference between mental state and physical state evolution. While he views the latter as a deterministic process, he regards the first, i.e., mental state evolution, as a probabilistic process. In the SI, likewise, if distribution of renewal at $t_0$ is $|\Psi(t_0)|^2$, then Schrödinger equation *deterministically* gives us the distribution $|\Psi(t)|^2$. However, the system's renewals occur at every moment, at best, according to a statistical distribution, and accordingly, we can attribute a probability to the ensemble.[1]

[1] Since these renewals are happening at every moment so they are superluminal but relativity governs on ensembles of them. Compare this with what Bohm said: at the underlying level there could be nonlocal behavior but at the ordinary level, we experience covariant behavior (Bohm & Hiley 1993:286).

In the Sigma model, the physical state of the flowing quantum object always evolves in the usual linear way; however, to have a complete theory, we also need to specify an *auxiliary dynamic*. Perhaps the simplest dynamics would be one like the configuration dynamics in Bell's Everett theory, where the probability of observer's current mental state is fully determined by the current physical state alone and is always equal to the norm-squared of the component of the universal state that describes observer as currently believing the corresponding determinate outcome.[2] The Sigma interpretation, with completely random stage dynamics, as discussed above, would correspond to something like Bell's Everett theory. The continuous and random renewal of temporal parts of the flowing system causes the observer's mental state to jump randomly, regardless of its last state.

In the Sigma model, the observer's mind, as in the single-mind or many-minds theory, is not a quantum mechanical system; it is never in superposition. This is what is meant by saying that it is non-physical. The time evolution of the observer's mind in the Sigma interpretation is, just as in the many-minds or single-mind theory, probabilistic. Furthermore, although the evolution of individual stages or renewals is probabilistic, the evolution of the set of stages associated with a flowing quantum object is deterministic, as the evolution of the measurement process is deterministic. We can infer the density of the stages in various states from the final state.

Evolution of flowing systems, in the Sigma interpretation, is both causal and stochastic. From the observer's point of view, it is stochastic, but the probabilistic feature is *not inherent* in nature; instead, it originates from the specific statistical distribution of the renewal of stages in the flowing system. This indeterminism is not necessarily a sign of intrinsic chances in nature. Rather, it means objects involve continuous renewal with a specific distribution. Of course, we should pursue a plausible account for this distribution. This is close to the Bohm's interpretation. We have an *implicit* order, which is the source of *explicit* order. The observer, who does not see all dimensions, only sees a determinate outcome, and, at this level, he could describe all that he sees but could not give an intelligible account for it.

The interpretation that takes the statistical predictions of quantum mechanics as the result of our ignorance —the so-called ignorance interpretation —is not consistent with the real superposition states and empirical outcomes of quantum mechanics (Auletta 2000: 107-115). In the SI, likewise, we could not adhere to the ignorant interpretation of probabilities since that specific distribution is part and parcel of the *flowing* system. Despite this fact, pursuing hidden variables for individual events is not only warranted but also recommended. Bohm, in his later works, took the stochastic feature irreducible, *not* in the sense that there is no causality. There

[1] In the Sigma interpretation, contrary to Everett's theory, *the whole* system does not exist *at a moment*. It has temporal parts and its whole reality can be found at the whole lifetime of system.

[2] This is similar to the single-mind analogue of the evolution of position in Bell's Everett theory. See (Bell 1987:133) and (Barrett 1999:187)

could be hidden variables at deeper levels.[1] He maintained that "… the chaotic or stochastic character may be contingent and determined by conditions in domains not covered by the particular theory in question. For example, random variations in the trajectory of a Brownian particle may be partially or even totally determined by atomic motions at a deeper level. Ultimately, our overall worldview is neither absolutely deterministic *nor absolutely indeterministic*. Rather, it implies that these two extremes are abstractions, which constitute *different views or aspects of the overall set of appearances*. Which view is appropriate in a given case will depend both on the *unknown totality* and on *our particular mode of contact with it* (e.g. the kinds of experiments we are able to do)." (Bohm & Hiley 1993: 324; *italics are mine*).

So, we can and should still raise the question *of why* such a distribution governs these renewals. We need an independent justification, other than the adequacy of quantum mechanics, to account for this distribution, and this is not a problem specific to the SI alone; Bohm's and Everett's models also have such a postulate and should provide a plausible account for it.

## 6. Problem of Effect

Consider an electron with positive x-spin entering a device with a z-spin measuring device. The outgoing electron has a z-up spin or z-down spin. The repeatability condition requires if the first measurement outcome is up, then the result of the second measurement, which is *immediately* performed, also certainly and definitely ought to be up. This is known as the *problem of effect*.

These considerations make "plausible" (but of course, are not intended to 'prove') one of the postulates of standard quantum mechanics, which tells us how the state vector is affected by a measurement: A measurement of an observable generally causes a drastic, uncontrollable alteration in the state vector of the system; specifically, regardless of the form of the state vector just before the measurement, immediately after the measurement it will coincide with the eigenvector corresponding to the eigenvalue obtained in the measurement.

Ballentine criticised Dirac (1958, 36) for taking such a condition seriously. He maintained that expecting the second measurement of the same observable, immediately after the first measurement, to yield the same result as the first measurement.

> "… is true at most for the very special class of measurements that do not change the quantity being measured. A statement of such a limited applicability is hardly suitable to play any fundamental role in foundations of quantum theory. In fact, this argument is based on the implicit (and incorrect) assumption that *measurement* is equivalent to *state preparation* … . For example, a Polaroid filter placed in the path of a photon beam constitutes a *state preparation* with respect to the polarisation of any transmitted photons. A second Polaroid at the same angle has no further effect. But neither of these processes constitutes a *measurement*. To measure the polarisation of a photon one must also detect whether or not

[1] This is similar to Ballentine's statistical interpretation. See (Ballentine 1970:380), (Auletta 2000:106).

> the photon was transmitted through the Polaroid filter. Since the detector will absorb the photon, no second measurement is possible." (Ballentine 1970:369).

This argument, however, seems to be inconclusive. Consider an x-spin measurement on a y-spin up electron. In any theory, this measurement will change the quantity being measured: before the measurement, in the standard interpretation, the electron does not have an x-spin. But *immediately* repeating the measurement will certainly give the same result. The absorption of the particle in certain measurement techniques is incidental, and one could imagine ways of carrying out the measurement that does not do this. The predictions of the theory for the measurement outcome would not be changed. There are classes of measurements associated with projection operators, for which repeating the measurement immediately will certainly give the same result. There will be correlations between successive results, which is another way to have the problem of effect. So, as Maudlin said:

> "The result of a measurement therefore, has predictive power for the future: after the first measurement is completed, we are in a position to know more about the outcome of the second than we could before the first measurement was made. Any theory which seeks to replicate the empirical content of the traditional theory should have this feature." (1995:13).

Similarly, Bell, in a critical assessment of Everett's theory, pointed out that Everett's model, like Bohm's, should suggest a dynamics that preserves reliability and yields the previous result in repeated measurements (Bell 1964:133, 135-136; Barrett 1999:80, 124-126, 185).

But what about the Sigma model? What does the observer *actually* measure? According to the Sigma model, the probability of the actual outcome ending up associated with a particular term in the quantum-mechanical state is completely determined by the current quantum-mechanical state and is independent of past outcomes. It is given by the square of the coefficients of the terms when the wave function is written in the related basis.

The system, in this interpretation, is in a state of perpetual renewal. These renewals could occur in a manner completely similar to the previous one, and from the empirical observation and observer's standpoint, it is as if the same value has been obtained. This situation, however, cannot be sufficient to solve the problem of effect. It would typically be possible that the observer's immediate measurement would, in fact, lead to wildly different outcomes. This situation is similar to Bell's Everrett(?) theory. As Bell put it:

> In our interpretation of the Everett theory there is no association of the particular present with any particular past. And the essential claim is that this does not matter at all. For we have no access to the past. We have only our 'memories' and 'records'. But these memories and records are in fact present phenomena The theory should account for the present correlations between these present phenomena. And in this respect we have seen it to agree with ordinary quantum mechanics, in so far as the latter is unambiguous. (Bell 1987: 135-6)

Bell objected that "if such a theory were taken seriously, it would hardly be possible to take anything else seriously" (1987: 136; see also 1987: 98). The same is true for the SI. While, generally and typically, it is in agreement with the statistics of standard theory, there is no guarantee that a correlation will exist between the present subjective result and later or former mental results.[1] Accordingly, the Sigma theory, just as Everett-like theories, requires an explicit rule to connect states at different times (Barrett 1999:197-198), and without such an explicit auxiliary dynamics, the Sigma theory seems to be incomplete. This auxiliary dynamics, which can be obtained as Bohm's approach for the particle path equation, describes how the determinate local stages (temporal parts of the quantum object) evolve, while the universal wave function evolves in its usual linear way. [2]

## 7. Conclusion

Every rational theory, whether scientific or philosophical, is rational in so far as it tries to *solve certain problems*.[3] Metaphysics provides the framework for science, but not in the sense of Descartes' or Kant's improved version of it, which, for good reasons, has been universally rejected. The interaction between physics and metaphysics is reciprocal, operating within a critical framework. Although metaphysical doctrines are generally less falsifiable than scientific theories, they remain open to critique regarding their internal coherence and their efficacy in addressing the problems they aim to resolve. Furthermore, when two competing metaphysical theories provide divergent interpretations of an established body of empirical evidence, the theory that aligns with the currently best-supported scientific explanation tends to prevail. Conversely, the metaphysics underlying the incompatible theory diminishes in explanatory utility and is eventually discarded. In this way, some scientific issues retain substantive relevance for metaphysical inquiry.[4]

In this line of thought, we attempted to show that the Sigma interpretation, as a theoretical proposal for the measurement problem based on the four-dimensional metaphysics, is *as reasonable and rational as* rival theories to give an account of the measurement problem, by assessing its ability and capacity to solve the three measurement problems, i.e. the problem of outcome, the problem of statistics and the problem of effects.

SI is an attempt to show that, if we want to be *realist* about quantum theory, if we want to accept the entities and states of the theory as *real*, we need to suggest new conceptual schemes, but *neither* in an isolated way nor in an ad-hoc manner. It suggests, by treating superposition states as real, a new conceptual framework, or, in Bohm's words, a "new order" for our universe (Bohm 1980:175). Despite some similarities, SI, by adopting four-dimensionalism as its *metaphysics*, is sufficiently differentiated from other interpretations.

[1] It seems that all subjective interpretations, even von Neumann's own interpretation, which takes collapse as a subjective process, involves such a difficulty.

[2] Stage dynamics, in Sigma, is on a par with mental states in the many-minds interpretation, as Albert said "What's been said so far … doesn't amount to a completely general set of laws of the evolution of mental states; but laws like that can be cooked up, and they can be cooked up in such a way as to guarantee that everything I've said about them so far will be true."(1992:129)

[3] This doctrine has been mainly developed by Popper. See for example, Popper (1958).

[4] See Agassi (1964)

Sigma, like Everett's model and Bohm's, has an advantage over the standard interpretation. All of them, unlike standard interpretation, are completely explicit about what really exists and how they evolve. There is no shifty division between the macro and micro worlds. There are also some additional advantages: First, unlike the bare theory, Sigma is in accord with our very deep conviction that mental states never superimpose. Nonetheless, it remains true to Everett's fundamental idea that linear dynamics govern the time evolution of the entire universe and every physical system: there is no need to postulate collapses, splits, or any other non-quantum mechanical physical phenomena. The reason that 'physical' is emphasised here is that Sigma, like many-minds theory, supposes the existence of a nonphysical entity, namely the observer's mind, whose states are not determined by the global quantum-mechanical state of the flowing object, and it makes possible the self-consciousness of the observer to his trans-temporal identity. Sigma theory also provides a particularly natural way to make sense of the claim that, at the end of a typical measurement, while in one sense, there is only one physical observer and one physical system, in another sense there are many *potentialities* with mutually incompatible experiences. But, contrary to many-minds theory, there is no need, in the SI, to consider *continuous infinite minds* for the observer, and to encounter difficulties in interpreting what it means to have many minds and how it is possible to have these minds, while all belongs to an *individual* observer, *without* any connection between them. There is no more than one mind in the SI, but this mind encounters continuous and infinite renewals of the system. Moreover, because of the *reality* and genuine nature of superposition states, Sigma model, has no problem concerning *personal identity*.

Everett gives no explanation for the total lack of influence of one branch on another. He, it seems, takes it for granted to explain why one would not feel the branching process. The same is true for *unobservability* of the splitting of universe in many-worlds theory. In the Sigma theory, however, there is an account for it: none of the two temporal parts of a *flowing* object (a four-dimensional object), as a *flowing continuum*, like time, exists together. Moreover, taking this model seriously leads us to the conclusion that quantum theory reveals yet another limitation of our perceptual experience, namely our inability to perceive the *unsharp reality*.[1]

Despite these advantages, the four-dimensionalism proposed in the SI may still provoke "incredulous stares"! We cannot consider the type of relation of mental state to physical state that Sigma theory provides as *supervenience* in a usual sense.[2] If we accept mental supervenience, we might want the mental state to supervene on physical state. However, based on the Sigma theory, the observer is associated with an infinite set of stages that are most likely associated with wildly contradictory beliefs and whose mental states the observer cannot know at any single moment. Although, the fact that the observer is associated with a continuous infinity of stages is at least counter-intuitive, this theory is not so counter-intuitive as many-minds which associates each observer with an *infinity of minds*. However, although we have mentioned some advantages and positive arguments in favour of SI, there is no claim of advantage at this stage. The claim of the paper, as mentioned in the abstract, is modest: the SI is as *rationally plausible as* its rivals and deserves to be criticised.

[1] It is a virtue for a scientific or philosophical theory to specify, in some ways, the *limitations* of our knowledge. Relativity shows our limitations in accessing information outside of the light cone. Godel's theorem specifies limitations of certain axiomatic systems. Kant's theory of knowledge attempted to show limitations of reasonable knowledge. All of these theories, regardless of their plausibility, specify, in some ways, limitations of our knowledge.

[2] The Sigma interpretation is *not* unique concerning *supervenience*. We encounter similar situation in von Neumann's interpretation (Becker 2004:127), and Bohm's theory (Pylkkanen 2007:192). It is also worth mentioning that *mind-body dualism* is not specific to SI. The many-minds and von Neumann Interpretation have the same presupposition of dualism. Moreover, since a clear account of mind/body interaction concerns the general mind/body problem, so this is not a problem that this paper aims to tackle with.

Finally, as Sklar (1974:2) and Balashov (1999) put it, there is an interdependence between science and philosophy. It is not simply that one cannot do good philosophy without relying upon the results of scientific theorising. Although science constantly forces us to revise and refine our commonsense and intuition, the acceptance or rejection of particular scientific theories depends as much on the adoption of philosophical presuppositions as it does on the evidence of observation and experimentation. The measurement problem is sometimes portrayed as merely philosophical, or of no interest to physics proper. This, as Maudlin (1995) said, is quite untrue. Each model carries with itself an obligation, which demands the postulation of *new physics*.